\documentclass[twoside,twocolumn,9pt]{article}
\usepackage{extsizes}
\usepackage[super,sort&compress,comma]{natbib}
\usepackage[version=3]{mhchem}
\usepackage[left=1.5cm, right=1.5cm, top=1.785cm, bottom=2.0cm]{geometry}
\usepackage{balance}
\usepackage{times,mathptmx}
\usepackage{sectsty}
\usepackage{graphicx}
\usepackage{lastpage}
\usepackage[format=plain,justification=justified,singlelinecheck=false,font={stretch=1.125,small,sf},labelfont=bf,labelsep=space]{caption}
\usepackage{float}
\usepackage{fancyhdr}
\usepackage{fnpos}
\usepackage[english]{babel}
\usepackage{array}
\usepackage{droidsans}
\usepackage{charter}
\usepackage[T1]{fontenc}
\usepackage[usenames,dvipsnames]{xcolor}
\usepackage{setspace}
\usepackage[compact]{titlesec}

\usepackage{epstopdf}
\usepackage{bm}        
\definecolor{cream}{RGB}{222,217,201}

\begin{document}

\pagestyle{plain}
\thispagestyle{plain}{

\renewcommand{\headrulewidth}{0pt}
}

\makeFNbottom
\makeatletter
\renewcommand\LARGE{\@setfontsize\LARGE{15pt}{17}}
\renewcommand\Large{\@setfontsize\Large{12pt}{14}}
\renewcommand\large{\@setfontsize\large{10pt}{12}}
\renewcommand\footnotesize{\@setfontsize\footnotesize{7pt}{10}}
\makeatother

\renewcommand{\thefootnote}{\fnsymbol{footnote}}
\renewcommand\footnoterule{\vspace*{1pt}%
\color{cream}\hrule width 3.5in height 0.4pt \color{black}\vspace*{5pt}}
\setcounter{secnumdepth}{5}

\makeatletter
\renewcommand\@biblabel[1]{#1}
\renewcommand\@makefntext[1]%
{\noindent\makebox[0pt][r]{\@thefnmark\,}#1}
\makeatother
\renewcommand{\figurename}{\small{Fig.}~}
\sectionfont{\sffamily\Large}
\subsectionfont{\normalsize}
\subsubsectionfont{\bf}
\setstretch{1.125} 
\setlength{\skip\footins}{0.8cm}
\setlength{\footnotesep}{0.25cm}
\setlength{\jot}{10pt}
\titlespacing*{\section}{0pt}{4pt}{4pt}
\titlespacing*{\subsection}{0pt}{15pt}{1pt}

\fancyfoot[RO]{\footnotesize{\sffamily{1--\pageref{LastPage} ~\textbar  \hspace{2pt}\thepage}}}
\fancyfoot[LE]{\footnotesize{\sffamily{\thepage~\textbar\hspace{3.45cm} 1--\pageref{LastPage}}}}
\fancyhead{}
\renewcommand{\headrulewidth}{0pt}
\renewcommand{\footrulewidth}{0pt}
\setlength{\arrayrulewidth}{1pt}
\setlength{\columnsep}{6.5mm}
\setlength\bibsep{1pt}

\makeatletter
\newlength{\figrulesep}
\setlength{\figrulesep}{0.5\textfloatsep}

\newcommand{\topfigrule}{\vspace*{-1pt}%
\noindent{\color{cream}\rule[-\figrulesep]{\columnwidth}{1.5pt}} }

\newcommand{\botfigrule}{\vspace*{-2pt}%
\noindent{\color{cream}\rule[\figrulesep]{\columnwidth}{1.5pt}} }

\newcommand{\dblfigrule}{\vspace*{-1pt}%
\noindent{\color{cream}\rule[-\figrulesep]{\textwidth}{1.5pt}} }

\makeatother

\twocolumn[
  \begin{@twocolumnfalse}
\vspace{3cm}
\sffamily
\vspace*{-3.0cm}
\begin{tabular}{m{4.5cm} p{13.5cm} }
\vspace*{-5.0cm}
\includegraphics{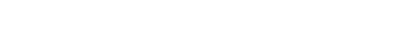} & \noindent\LARGE{\textbf{Product lambda-doublet ratios for the O($^3$P) + D$_2$ reaction: A
mechanistic imprint.}} \\
\vspace{0.3cm} & \vspace{0.3cm} \\

 & \noindent\large{P. G. Jambrina,\textit{$^{a}$} A. Zanchet,\textit{$^{a}$} J. Aldegunde,\textit{$^{b}$} M. Brouard,\textit{$^{c}$}   and F. J. Aoiz$^{\ast}$\textit{$^{a}$}} \\
   & \noindent\normalsize{ $^{a}$ Departamento de Qu\'{\i}mica F\'{\i}sica I, Universidad Complutense de Madrid, Spain. } \\
   & \noindent\normalsize{ $^{b}$ Departamento de Qu\'{\i}mica F\'{\i}sica , Universidad de Salamanca, Spain.} \\
   & \noindent\normalsize{ $^{c}$ The Department of Chemistry, University of Oxford, United Kingdom } \\ 

\includegraphics{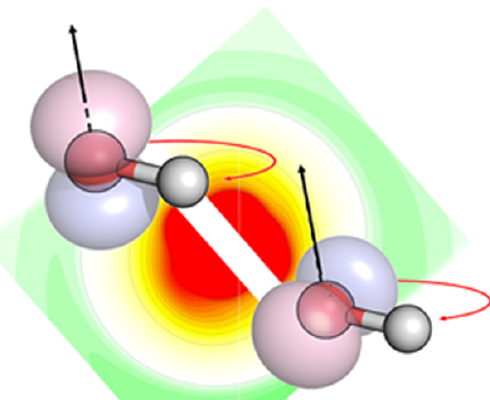} &\vspace*{-1.5cm} \noindent\normalsize{In the last decade, the development of theoretical methods have allowed
chemists to reproduce and explain almost all of the experimental data
associated with elementary atom plus diatom collisions. However, there are
still a few examples where theory cannot account yet for experimental results.
This is the case for the preferential population of one of the
$\Lambda$-doublet states produced by chemical reactions. In particular, recent
measurements of the OD($^2\Pi$) product of the O($^3$P) + D$_2$ reaction have
shown a clear preference for the $\Pi(A')$ $\Lambda$-doublet states, in
apparent contradiction with {\em ab initio} calculations, which predict a
larger reactivity on the $A''$ potential energy surface. Here we present a
method to calculate the $\Lambda$-doublet ratio when concurrent potential
energy surfaces participate in the reaction. It accounts for the experimental
$\Lambda$-doublet populations via explicit consideration of the stereodynamics
of the process. Furthermore, our results demonstrate that the propensity of the
$\Pi(A')$ state is a consequence of the different mechanisms of the reaction on
the two concurrent potential energy surfaces.}

\end{tabular}

 \end{@twocolumnfalse} \vspace{0.6cm}

  ]

\renewcommand*\rmdefault{bch}\normalfont\upshape
\rmfamily
\section*{}
\vspace{-1cm}

\footnotetext{*E-mail: aoiz@quim.ucm.es}

%




\section{Introduction}

Chemists are keen to describe chemical reactions in terms of the motion of
billiard balls on a more or less complex quantum electronic landscape, the
Potential Energy Surface (PES). However, this picture is not always valid and
quite often, several PESs have to be considered, giving rise to non-adiabatic
effects that may have a decisive influence on the dynamics. When this is the
case, it is not possible to disentangle experimentally the contribution of each
of the competing surfaces, answering the question of which of them is more/less
reactive, and why. The presence of multiple PESs correlating reactants and
products leads to open shell molecules in which the rotational levels are split
in spin-orbit states, and, in turn, each of them in two nearly degenerate
$\Lambda$-doublet levels that can be spectroscopically resolved due to
different selection rules. In spite of the tiny energy difference between the
$\Lambda$-doublet pair of states, a clear preference towards one of them is
observed in many chemical reactions
\cite{MLML:JCP78,faraday86,BZRZ:CPL89,HCW:CPL00,LZMM:NC13, LZMM:JACS14}. As
pointed out by several authors, \cite{Adresen:JCP1985, Alexander:JCP84,
SEAHB:PRL85,B:JCP86, AS:BOOK87, DAL:JCP89,Alexander1988,
BZ:CPL90,D:JCP95,A:NC13} the $\Lambda$-doublet population acts as a fingerprint
to unravel the symmetries of the surfaces involved in the process so that the
propensity for one of the manifolds reflects the competing reactivity of
concurrent PESs and addresses the question of where the electrons go when the
reaction takes place \cite{BZ:CPL90,A:NC13}.   However, a general, clear-cut
relationship between them has not yet been demonstrated.

Collisions leading to NO$(^{2}\Pi)$ and OH$(^{2}\Pi)$ are prototypical for the
study of $\Lambda$-doublet propensities. Recent experiments by Minton,
McKendrick and coworkers \cite{LZMM:NC13,LZMM:JACS14} have determined the
OD($X^2\Pi$) state-to-state $\Lambda$-doublet population ratios for
O($^3$P)+D$_2$ collisions. Regardless of the collision energy and final
vibrational state, they consistently found a significantly larger population of
the $\Pi(A')$ $\Lambda$-doublet state compared to the $\Pi(A'')$ one, where the
labelling of the states refers to the location of the singly occupied orbital
in the rotation plane of the diatom, $\Pi(A')$, or perpendicular to it,
$\Pi(A'')$, in the  limit of high products rotational states
$j'$.\cite{AR:JCP85, Alexander1988,D:JCP95} This result seems to contradict the
theoretical results, which would predict a preference for $\Pi(A'')$ under the
assumption that for the two concurrent PESs, of symmetry $^{3}A'$ and
$^{3}A''$, collisions on the first one will only form the $\Pi(A')$
$\Lambda$-doublet state and vice-versa. This simple assignment is supported by
the rationale that for direct, sudden collisions, the products ``remember'' the
collision conditions and, thus, there should be a close relationship between
both symmetries. It should be pointed out that a general procedure to connect
the reactivity on concurrent surfaces with the $\Lambda$-doublet population has
not been achieved. In what follows, we  present a method that connects the
reactivity on the $A'$ and $A''$ PESs with the populations of the respective
$\Lambda$-doublet states, through the explicit consideration of the reaction
stereodynamics. As will be shown, this method is capable to reproduce and to
explain the origin of the experimental $\Lambda$-doublet propensities.

This article is organize as follows: In the theory section we will present the
method (the interested reader is referred to the Appendix for a more detailed
presentation). In the Results and Discussion section, the OD($X^2\Pi$)
$\Lambda$-doublet population ratios for O($^3$P)+D$_2$ reaction have been
calculated and compared with the experimental results. The present theory also
allows us to connect the predicted $\Lambda$-doublet propensities with the
reaction mechanism. Finally, the main conclusions will be summarized.

\section{Theory}

We will start by invoking conservation of the reactive flux, which implies that
the population of the two $\Lambda$-doublet states and the cross sections on
the $A'$ and $A''$ PESs are related by
\begin{eqnarray} \label{eqrelationship1}
  \sigma_{v'j'}(\Pi(A')) &=& W_{A'} \, \sigma_{v'j'}{(A')}  + (1-W_{A''}) \,\sigma_{v'j'}{(A'')} \\
  \sigma_{v'j'}(\Pi(A'')) &=&  (1-W_{A'}) \, \sigma_{v'j'}{(A')} + W_{A''} \, \sigma_{v'j'}{(A'')} \,, \label{eqrelationship2}
\end{eqnarray}
where $\sigma_{v'j'}{(A')}$ and $\sigma_{v'j'}{(A'')}$ are the rovibrational
state resolved cross sections on the two respective PESs and $W_{A'}$ and
$W_{A''}$  represent the ``correction factors'' to obtain the $\Lambda$-doublet
cross sections for a given $v', j'$ rovibrational state. As commented on above,
in the sudden limit, the flux ending on the $A'$ PES is assigned to the
$\Pi(A')$ state and {\em vice versa}, which is equivalent to setting
$W_{A'}=1$, and $W_{A''}=1$.

For a given nuclear geometry, the weights $W_{A'}$ and $W_{A''}$ are the square
of the coefficients that define the expansion of the D--OD asymptotic
electronic wavefunctions in terms of the $\Lambda$-doublet molecular
wavefunctions $\varphi[\Pi(A')]$ and $\varphi[\Pi(A'')]$, \cite{BZ:CPL90}
\begin{eqnarray}\label{wf1}
  \psi_{A'}  &=& a^{A'}_1 \,\varphi[\Pi(A')] +  a^{A'}_2 \,\varphi[\Pi(A'')] \\
   \psi_{A''}  &=& a^{A''}_1 \,\varphi[\Pi(A')] +  a^{A''}_2
   \,\varphi[\Pi(A'')] \label{wf2}
\end{eqnarray}
These coefficients are related to the dihedral angle between the three-atom
plane and the OD molecular plane.\cite{BZ:CPL90} This angle connects the
symmetry of the PES to that of the $\Lambda$-doublet state, and, in the high
$j'$ limit, can be identified with $\theta_{{\bm j}' {\bm u}}$, the angle
between the rotational angular momentum, $\bm{j'}$, perpendicular to the OD
rotation plane and the vector $\bm u$ perpendicular to the three-atom plane
\cite{PHMKAB:JCP15} (see Fig. \ref{frame} in the Appendix). For the $A'$ PES the singly
occupied orbital  lies in the triatomic plane and, hence, $a^{A'}_1 = \cos
\theta_{{\bm j}' {\bm u}}$, and $a^{A'}_2 = \sin \theta_{{\bm j}' {\bm u}}$.
Conversely, for the $A''$ PES, the orbital lies perpendicular to the triatomic
plane, leading to $a^{A''}_1 = -\sin \theta_{{\bm j}' {\bm u}}$, and $a^{A''}_2
= \cos \theta_{{\bm j}' {\bm u}}$.

To obtain the weights $W_{A'}$ and $W_{A''}$, one just needs to average $\cos^2
\theta_{{\bm j}' {\bm u}}$ over one rotational period for calculations on the
$A'$ and $A''$ PES respectively. This is straightforward in the quasiclassical
trajectories (QCT) framework,\cite{PHMKAB:JCP15} where $\theta_{{\bm j}' {\bm
u}}$ can be computed at every step of the trajectory. In a pure quantum
mechanical (QM) context the equivalent magnitude would be $\langle {j'}_{\bm
u}^2 \rangle / (j (j+1))$, where ${j'}_{\bm u}^2$ is the projection of the
rotational angular momentum along the ${\bm u}$ vector.

A crucial finding, that can be demonstrated using either QCT or QM arguments
(see Appendix), is that $\langle \cos^2\theta_{{\bm j}' {\bm
u}} \rangle_{ \rm rot}$, the average value of the square angle cosine is
related to the helicity, $\Omega'$, the projection of $\bm{j'}$ on the products
recoil direction ($\bm{k'}$), through the expression:
\begin{equation}\label{cos2eq1}
\langle \cos^2\theta_{{\bm j}' {\bm u}}  \rangle_{ \rm rot}   = 1 -  \left|\frac{\Omega'\,^2}{j'
(j'+1)}\right|^{1/2} \,,
\end{equation}
Equation \eqref{cos2eq1} has very important implications: $W_{A'}$ and
$W_{A''}$ for a given rovibrational state depend only on the distribution of
the helicities and, in general, will differ because such distributions reflect
the mechanisms on the concurrent PESs, that can be different. This means that
equations \eqref{eqrelationship1} to \eqref{cos2eq1} can be used to: i)
determine $\Lambda$-doublet populations also in a purely QM context, for which
$\Omega'$ is well defined, and ii) relate the $\Lambda$-doublet populations
with the reaction mechanism (see below).

The average value of $|\Omega'\,|^2$ can be determined from the product
rotational alignment moment, $a^{(2)}_0(j')$, which contains the essential
information about the alignment of $\bm j'$ with respect to the product recoil
velocity, and is given by \cite{miranda:1999jcp}
\begin{equation} \label{a20}
a^{(2)}_0(j')= \frac{\displaystyle \sum_{\Omega'} \sigma_{v'j'}(\Omega')~
\langle j' \Omega', 2 0 | j' \Omega' \rangle} {\sigma_{v'j'}} =
\frac{\displaystyle \sum_{\Omega'} \sigma_{v'j'}(\Omega')~ [3 \Omega'^2
-j'(j'+1)]} {2 \, C\,\sigma_{v'j'}} \,,
\end{equation}
where $\sigma_{v'j'}(\Omega')$ is the cross section resolved in $(v', j',
\Omega')$, $\langle :\,\, :, 2 0 | : \,\, : \rangle$ is the Clebsch-Gordan
coefficient.  $C = \left[j' (j'+1) (j'-1/2) (j'+3/2) \right]^{1/2}$, which for
high enough $j'$ is $\approx j'(j'+1)$. The average value of $\Omega'$ for a
given $j'$ is
\begin{equation}
\langle \Omega'\,^2 \rangle = \frac{\displaystyle  \sum_{\Omega'} \sigma_{v'j'}(\Omega')~ \Omega'\,^2  }{\sigma_{v'j'}} =
\frac{2 C {\ }a^{(2)}_0(j')}{3 } + \frac{j'(j'+1)}{3} \,,
\end{equation}
leading to the following expression for $W_{A'}$,
\begin{equation}\label{qmweight}
W_{A'} = 1- \left[\frac{\langle \Omega'^2 \rangle}{j' (j'+1)} \right]^{1/2}
\approx 1 - \left[ \frac{2}{3} a^{(2)}_0(j')  + \frac{1}{3}\right]^{1/2} \,,
\end{equation}
where the polarization moments, $a^{(2)}_0(j')$, have been calculated on the
$A'$ PES. Identical expressions hold for $W_{A''}$ when the $a^{(2)}_0$
alignment moment is calculated on the $A''$ PES are used.
Equation.~\eqref{qmweight} has one important consequence: {\em the
stereodynamics of the products - specifically, the $\bm{k'}$-$\bm{j'}$
correlation - relates the $\Lambda$-doublet populations to the reactivity on
the $A'$ and $A''$ PESs}. Classically, $a^{(2)}_0$ lies in the $[-1/2,1]$
range, although its QM limiting values depends on $j'$. Negative values of
$a^{(2)}_0$, close to its lower limit, correspond to $\bm{j}' \bot \bm{k}'$ and
$|\Omega'| \approx 0$, whilst positive values, close to one, imply that
$\bm{j}' || \bm{k}'$ and $|\Omega'| \approx j'$. According to
eqn~\eqref{qmweight}, weight factors close to zero are associated with
$a^{(2)}_0\approx$1; that is, products on the $A'$ PES would appear as the $\Pi
(A'')$ $\Lambda$-doublet state and {\em vice versa}. When
$a^{(2)}_0\approx$--1/2, the weight factor tends to 1 and products on the $A'$
PES would correspond to the $\Pi (A')$ $\Lambda$-doublet state.

\section{Results and Discussion} \label{results}

\begin{figure*}
\centering
  \includegraphics[width=0.87\linewidth]{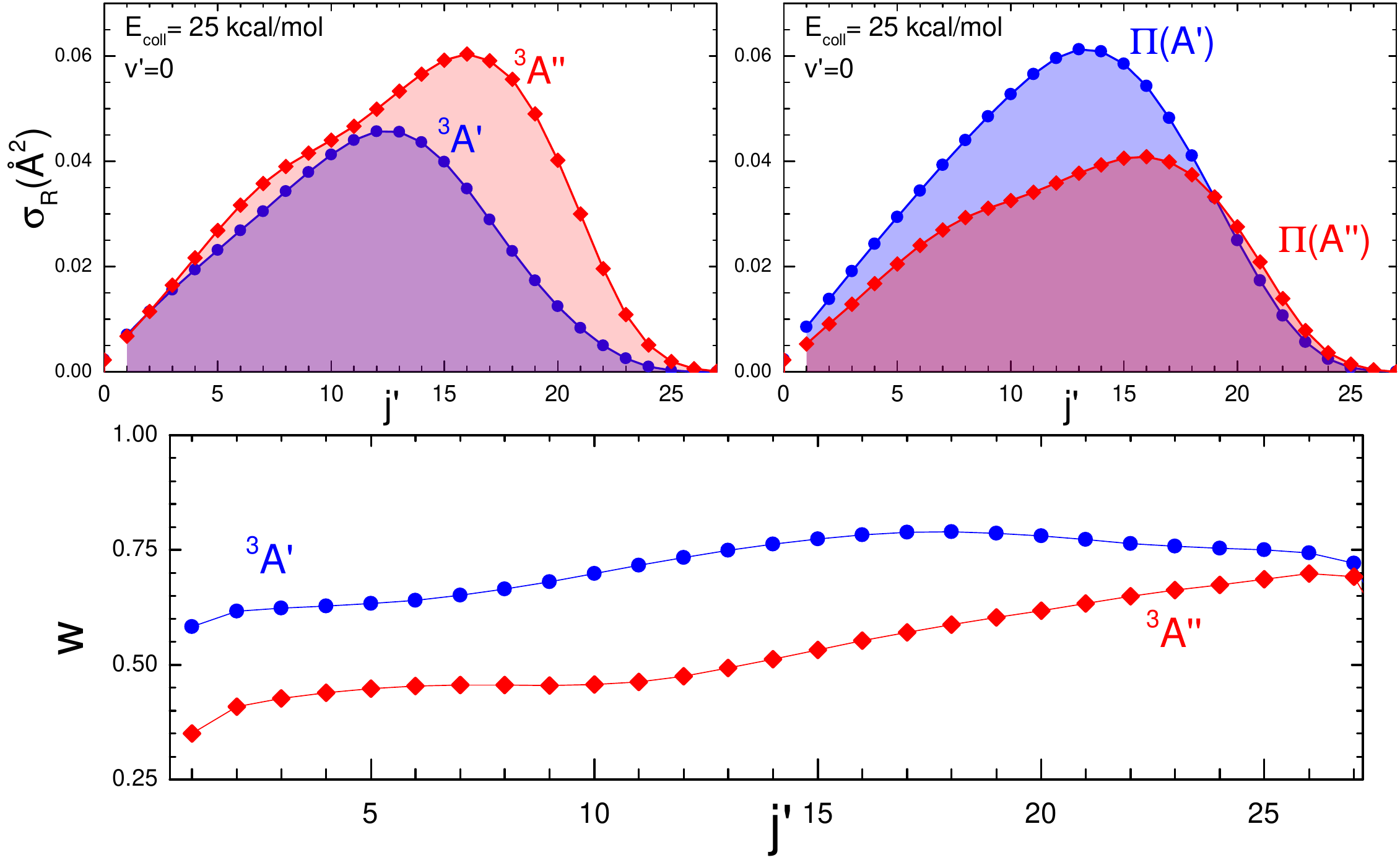}
\caption{QM weight factors to determine the $\Pi(A'')$ and $\Pi(A')$
populations from the integral cross section of the $^3A'$ and $^3A''$ PES. The
top left panel displays the reactive cross sections calculated on the $A'$ and
$A''$ PES, which, in the sudden limit, would represent the $\Pi(A')$ and
$\Pi(A'')$ state resolved cross sections. The top right panel shows the
reactive cross sections calculated for the two $\Lambda$-doublet  levels once
respective weights have been incorporated. The weights are shown in the bottom
panel. The data were obtained from the QM reaction cross sections for the
O($^3P$) + D$_2$ reaction at $E_{\rm coll} =25$\,kcal\,mol$^{-1}$.}
\label{fig1}
\end{figure*}

Aiming to test the method and to try to reproduce the experimental results of
Minton, McKendrick and coworkers \cite{LZMM:NC13,LZMM:JACS14}, we carried out
adiabatic time-independent QM and QCT calculations following the procedures described in Refs. \citenum{ABC,AHS:JCP92,ABH:JCSFAR98} using a  new  set of $^3A'$
and $^3A''$ PESs  (see method section for further details).

The adiabatic QM state-to-state reactive cross sections for the O($^3P$) +
D$_2$ reaction at $E_{\rm coll}= 25$\,kcal\,mol$^{-1}$, one of the energies of
the experiments carried out by Minton and coworkers\cite{LZMM:NC13}, are
represented on the top left panel of Fig.~\ref{fig1}. These results show that,
whilst for low OD($v' =0$, $j'$) rovibrational states the $A'$ PES is as
reactive as the $A''$ one, for $j'>12$ the integral cross sections (ICS) on the
$A''$ PES are considerably larger than those on the $A'$ PES. This is not
surprising as, although both PESs have the same barrier height,  the potential
energy raises faster with the bending angle for the $A'$ electronic state than
for the $A''$ state\cite{RWKW:JPCA00}, {\em i.e.}, the ``cone of acceptance''
is larger on the $A''$ PES. The inclusion of non-adiabatic couplings in the
dynamics does not change this picture, as  trajectory surface hopping
\cite{HS:JCP00} and non-adiabatic QM calculations \cite{HZ:JCC11,Z:JCP13} also
indicate larger reactivities on the $A''$ PES.

\begin{figure}
\centering
  \includegraphics[width=1.00\linewidth]{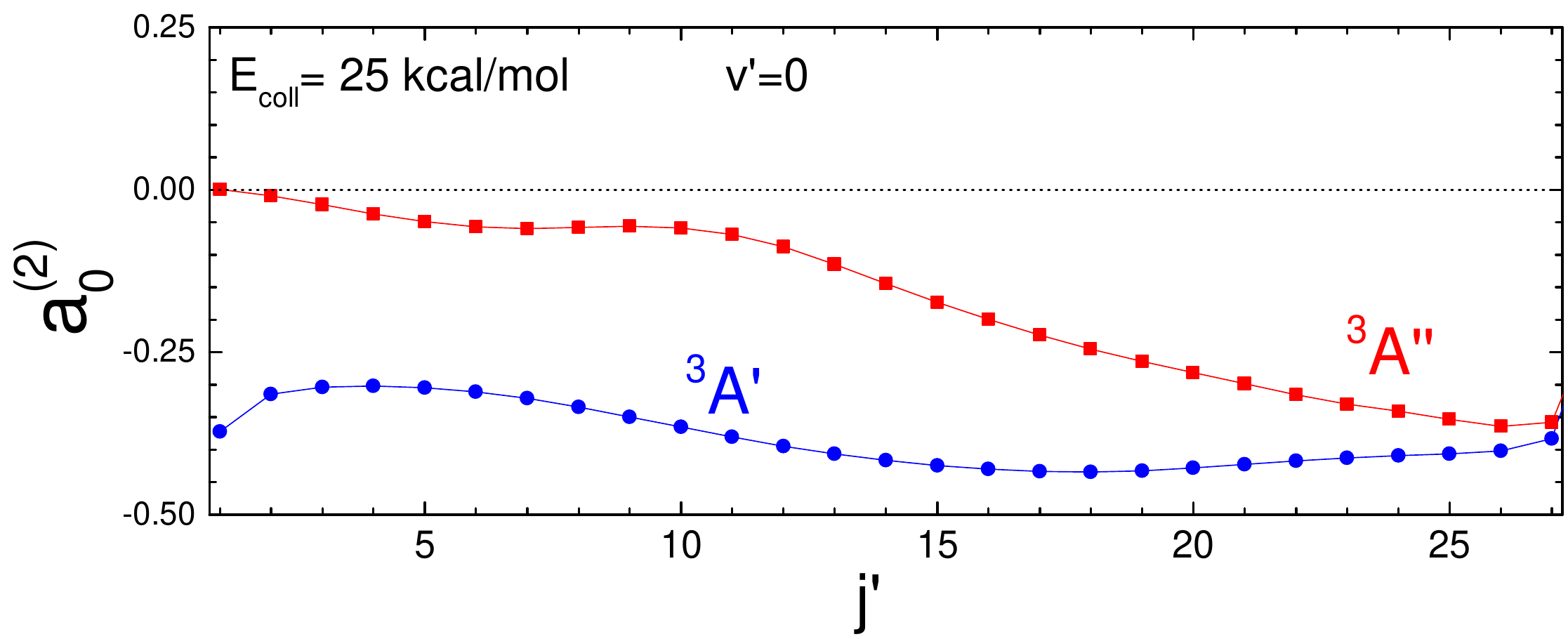}
\caption{Product QM alignment parameter, $a^{(2)}_0$, referenced to  $\bm k'$
(which defines the $z$ axis). The alignment moment $a^{(2)}_0$ is given by the
average value $C^{-1}\langle 3\hat{j'}_z^{2} -\hat{j'}^2\rangle/2 $, where
$\hat{j'}^2$ and $\hat{j'}_z$ are the rotational angular momentum operators and
$C$ is the constant that appears in eqn~\eqref{a20}. Calculations are presented
for O($^3$P) + D$_2$($v=0$, $j=0$) $\to$ OD($v'=0$, $j'$) + D at $E_{\rm coll}
=25$\,kcal\,mol$^{-1}$ on the $A'$ and $A''$ PESs. Whilst the product
rotational angular momentum, $\bm j'$, on the $A'$ PES is strongly polarized
perpendicular to $\bm k'$ (negative $a^{(2)}_0$ values), for the $A''$ PES the
distribution of $\bm j'$ is largely isotropic for low $j'$.} \label{fig5}
\end{figure}

Before discussing the other panels of Fig.~\ref{fig1}, it is pertinent to
inspect the integral alignment moments, $a^{(2)}_0$, which are shown in
Fig.~\ref{fig5} as a function of the rotational state for $v'=0$. The
differences between the values and the trends in $a^{(2)}_0(j')$ on the two
PESs are conspicuous, and indicate that very different stereodynamics are at
play on the two surfaces. For reaction on the $A'$ PES, $\bm j'$ is strongly
polarized perpendicular to the recoil direction, $\bm k'$, for essentially all
$j'$ states, and in some instances ($j'=15$-17) the $a^{(2)}_0$ values are very
close to the limiting negative value. In stark contrast, on the $A''$ PES, $\bm
j'$ is almost unpolarized for $j' \le$15, with small $a^{(2)}_0$ values close
to the isotropic limit, $a^{(2)}_0=0$. With increasing $j'$ above 15,
$a^{(2)}_0$ becomes gradually more negative approaching the values found on the
$A'$ PES.

Inserting the values of the alignment moments calculated on both PESs into
eqn~\eqref{qmweight} yields the weight factors, $W_{A'}$ and $W_{A''}$. They
are shown in the bottom panel Fig.~\ref{fig1} for $v'=0$ at $E_{\rm
coll}=25$\,kcal\,mol$^{-1}$. As can be seen, for $j'< 15$, $W_{A''}< 0.5$,
which, in effect, means that more than 50\% of the reactivity on the $A''$ PES
is ``transferred'' to the $\Pi(A')$ $\Lambda$-doublet state. In contrast, as a
result of the consistently fairly negative values of the alignment parameters,
$W_{A'}$ is always $>0.6$ and in some cases is a large as 0.80. In consequence,
the relative ``transfer'' of reactivity from $A'$ to $\Pi(A'')$ is much less significant
than that found from $A''$ to $\Pi(A')$.

\begin{figure*}
\centering
\includegraphics[width=0.71\linewidth]{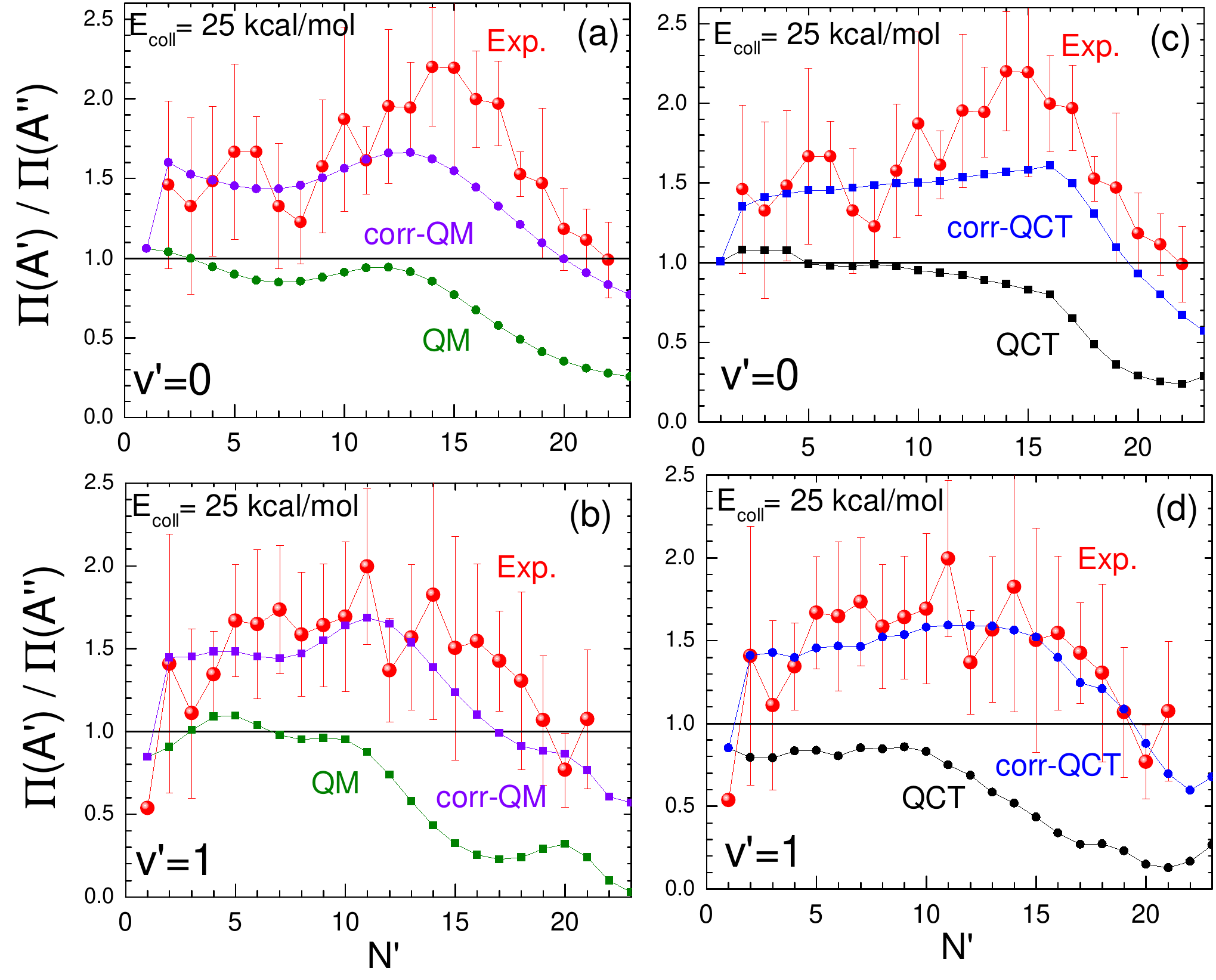}
\caption{Experimental (from Ref.\citenum{LZMM:NC13}), and the present QCT and
QM $\Lambda$-doublet population ratios for the O($^3P$) + D$_2$ reaction at
$E_{\rm coll} =25$\,kcal\,mol$^{-1}$. `QM' and `QCT' represent the ICSs
obtained on the $A'$ and $A''$ PES without any correction factor, whereas
`corr-QM' and `corr-QCT' are the $\Pi_{A'}/\Pi_{A''}$ ratios after making use
of the respective weight factors.}\label{fig2}
\end{figure*}
\begin{figure*}
\centering
  \includegraphics[width=0.71\linewidth]{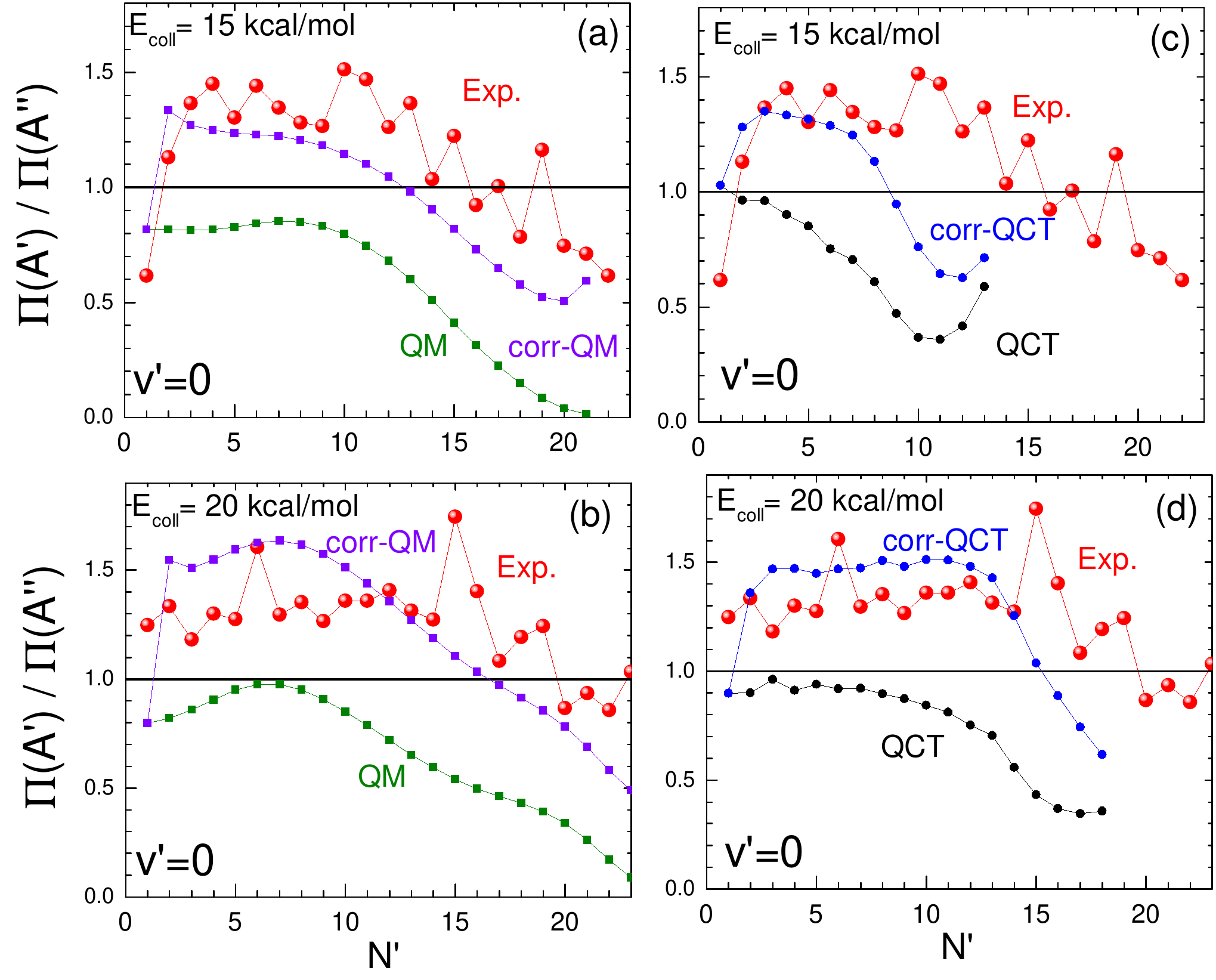}
\caption{Experimental, QCT, and QM $\Lambda$-doublet ratios for the  O($^3P$) +
D$_2$ reaction at $E_{\rm coll}=15$, 20\,kcal~mol$^{-1}$. The experimental
error bars were not reported at Ref.\citenum{LZMM:JACS14}.} \label{fig3}
\end{figure*}

Therefore, after correction, the relative population on the $\Pi(A')$ state is
significantly enhanced. The resulting $\Lambda$-doublet populations are
depicted in the right panel of Fig.~\ref{fig1}. Quite remarkably, the situation
is the reverse of that found for the  reactivity on the respective PESs: the
$\Pi(A')$, $\Lambda$-doublet state is considerably more populated than the
$\Pi(A'')$ state for low $j'$. In particular, for $j'=12$, $\sigma(\Pi_{A''}) =
(3/2) \times \sigma(\Pi_{A'})$. At higher $j'$ values ($j'>18$) the populations
of the two $\Lambda$-doublets are very similar.

In Fig.~\ref{fig2}, the experimental $\Lambda$-doublet population ratios
measured \cite{LZMM:NC13} at 25\,kcal\,mol$^{-1}$ are compared with the present
QM (left panels) and QCT (right panel) calculations for the $v'=0$, 1
manifolds. All the results are plotted against $N'=j'+1$, where $j'$ and $N'$
stand for the nuclear (closed-shell) and total (apart from spin) rotational
angular momentum, respectively. For each case, two series of results are shown:
(i) the ratio of the ICSs on the $A'$ and $A''$ (labeled as `QCT' and `QM')
where $W_{A'}$ and $W_{A''}$ are implicitly set to one, and (ii) the ratio of
the populations on the $\Pi(A')/\Pi(A'')$ using eqns. \eqref{eqrelationship1}
and  \eqref{eqrelationship2} with the correction factors included. For the
latter results (labeled as `corr-QM' and `corr-QCT'), the $W_{A'}$ and
$W_{A''}$ factors are calculated according to eqn~\eqref{qmweight}. It is
evident that the uncorrected QCT and QM results cannot account for the
experimental $\Lambda$-doublet ratios and, regardless of $j'$, predict larger
populations on the $\Pi(A'')$ state, in striking disagreement with the
experimental results. In contrast, the corrected results reproduce fairly well
the experimental values. In particular, the `corr-QM' results are within the
experimental error bars for most of the final states shown, particularly for
OD($v'=1$).

As shown in Fig.~\ref{fig3}, similar agreement between experimental
\cite{LZMM:JACS14} and theoretical results is obtained at $E_{\rm coll} =
20$\,kcal\,mol$^{-1}$. At even lower collision energies, $E_{\rm coll} =
15$\,kcal\,mol$^{-1}$, the agreement between the corrected QCT and experimental
results is not as good, probably because the collision energy is just above the
barrier. In fact, no trajectories were found for $N' >$13, whilst the QM and
experimental data populate up to $N'=21$. The corrected QM results remain in
good agreement with the experiments at this low collision energy. It is worth
noticing that while our corrected results predict quantitatively the
experimental $\Lambda$-doublet ratio regardless of the collision energy and
vibrational manifold studied,  uncorrected results fail to account
qualitatively the experimental measurements, predicting a preference towards
the $\Pi(A'')$ states.

\begin{figure}
\hspace*{-0.70cm}
  \includegraphics[width=1.08\linewidth]{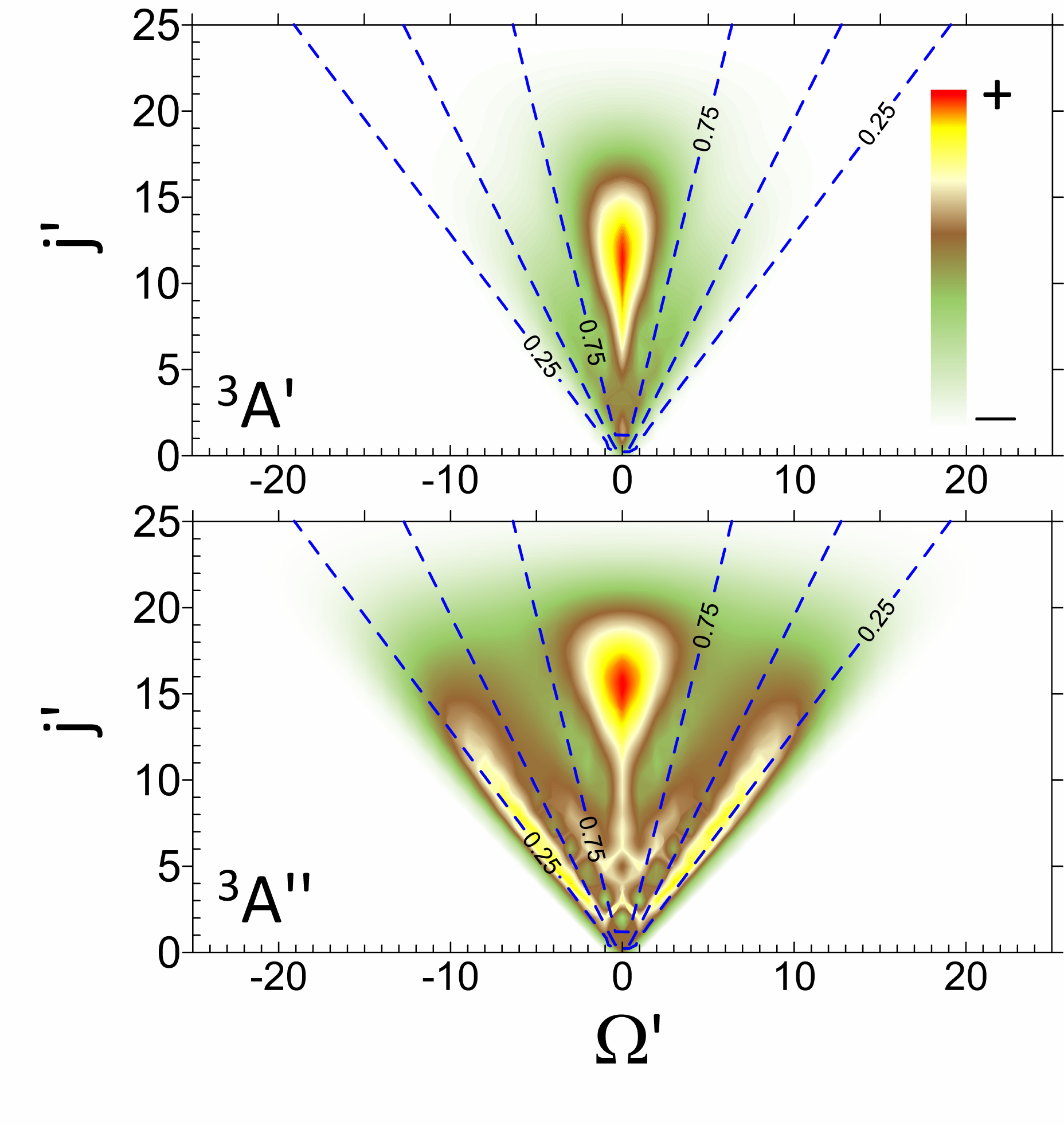}
\caption{Contour plots of the $\Omega'$-resolved QM cross sections,
$\sigma(v'=0, j',\Omega')$, at $E_{\rm coll} = 25$\,kcal\,mol$^{-1}$ calculated
on the $A'$ PES (top panel), and on the $A''$ PES (bottom panel) as a function
of both $|\Omega'|$ and $j'$. The contour lines indicate the values of $W_{A'}$
and $W_{A''}$ for a given combination of $j'$ and $|\Omega'|$.} \label{fig4}
\end{figure}

As already discussed, the way in which cross sections on the $A'$ and $A''$
PESs are combined to obtain the $\Pi(A')$ and $\Pi(A'')$ populations is
strictly related to the alignment of the product  rotational angular moment
with respect to the recoil direction. To show this effect more clearly, the
values of $\sigma(v'=0,j',\Omega')$ as a function of $\Omega'$ and $j'$ are
depicted as contour maps in Fig.~\ref{fig4} for the $A'$ and $A''$ PESs. The
differences between the respective contour maps are clear to see. The ICS for a
given $j'$ state includes the contribution from many $\Omega'$ values on the
$A''$ PES, whilst on the $A'$ the contribution is restricted to relatively few,
low $\Omega'$ values. Hence, this picture complements Fig.~\ref{fig5}. Negative
values of $a^{(2)}_0$ close to the limit imply that $\langle |\Omega'| \rangle$
is very small, nearly zero. If the contributions of higher $\Omega'$ values
becomes more significant, the alignment moment tends to be zero.

A more quantitative analysis can be carried out by relating the $\Omega'$
contributions to the weight factors that have been used to extract the
$\Lambda$-doublet populations. To this end, we have used iso-contour lines for
the different values of the weight factors. If in eqn~\eqref{qmweight},
$a^{(2)}_0$ is replaced with the $\langle j' \Omega', 2 0 | j' \Omega'
\rangle$, which is nothing but the $a^{(2)}_0$ for a pure ($j'$, $\Omega'$)
state, we can assign a single weight factor to every point on the $j'-\Omega'$
surface. On the $A'$ PES, most of the reactivity comes from low $\Omega'$
values ($\bm{j'} \bot \bm{k'}$), falling within the $W_{A'}>0.75$ limits shown
by the central dashed lines in Fig.~\ref{fig4}. In contrast, on the $A''$ PES
we find two different trends. For the highest $j'$ values ($j'> 15$) most of
the reactivity corresponds to low $\Omega'$, as for the $A'$ PES, although some
contributions from higher $\Omega'$ values can also be seen. However, with
decreasing $j'$, the low $\Omega'$ peak coexists with additional peaks
corresponding to $\Omega' \approx j'$ values ($\bm{j'} || \bm{k'}$), which
appear along the $W_{A''}\approx 0.25$ dashed lines. The averaging over these
two  contributions leads to a nearly isotropic alignment ($a^{(2)}_0 \sim 0$).
These contributions represent two distinct mechanisms: one coplanar which gives
rise to low $\Omega'$, and another one, that takes place only on the $A''$ PES,
which correlates with high $\Omega'$ states and for which the three-atom and OD
rotational plane tend to be orthogonal.


%

\section{Conclusions}

In spite of the tiny energy difference between the $\Lambda$-doublet pair of
states, a clear preference towards a particular $\Lambda$-doublet state is
observed for many chemical reactions. This intriguing fact has long puzzled
researchers and, actually, has been the subject of an ongoing discussion for
more than thirty years.

Throughout this article we have presented a new method that allows one to
extract the $\Lambda$-doublet populations from the cross sections on the $A'$
and $A''$ PESs.  It is shown that the transformation between the reactivities
on the $A'$ and $A''$ PESs and the $\Lambda$-doublet populations only requires
knowledge about the stereodynamics of the reaction, in particular of the
alignment of the products rotational angular momentum, $\bm{j}'$, along the
recoil direction, $\bm{k}'$. This procedure is in principle general and can be
used in combination with scattering data obtained using both QCT and QM
adiabatic and non-adiabatic methods.

This method has been applied to  the O($^3P$) + D$_2$ reaction, for which we
have carried out QCT and QM adiabatic calculations. Although couplings between
the concurrent PES has not been included, our method accounts quantitatively
for the experimental $\Lambda$-doublet populations obtained by Minton and
coworkers \cite{LZMM:NC13,LZMM:JACS14} that have thus far remained unexplained.
The analysis of the results has shown that the preference for the $\Pi(A')$
$\Lambda$-doublet state is due to the existence of an additional mechanism on
the $A''$ PES for which OD rotational plane tend to be orthogonal to the
three-atom plane. This mechanism has been traced back to the comparative
topographies of the $A'$ and $A''$ PES, the latter characterized by a broader
cone of acceptance.

\section{Computational Methods}

\subsection{Ab Initio Calculations:}

The PESs of the lowest $1^3A'$ and $1^3A''$ states were determined using 3500
{\em ab initio} points for each PES that were calculated using the MOLPRO suite
of programs. \cite{MOLPRO1,MOLPRO2} For both oxygen and hydrogen atoms, an
aug-cc-pV5Z basis set including {$spdfg$} basis functions was used. To obtain
an accurate and homogeneous description of the PESs, the state-average complete
active space (SA-CASSCF) method \cite{Werner-etal:85} was employed. The active
space considered consisted of 8 electrons distributed in 6 orbitals (2-6{$ a'$}
and 1{$a''$) in order to include all valence orbitals of oxygen and the 1{$s$
orbitals from both hydrogen atoms. The state-average orbitals and
multireference configurations obtained were then used to calculate both the
lowest $1^3A'$ and the lowest $1^3A''$ state energies with the internally
contracted multireference configuration interaction method (icMRCI), including
single and double excitations\cite{Werner-etal:88} and the Davidson correction.
\cite{Davidson:75}

The {\em ab initio} icMRCI+Q energies for the $1^3A'$ and $1^3A''$ electronic
states were fitted separately using the GFIT3C procedure introduced in
Refs.~\citenum{Aguado-etal:92,Aguado-etal:93,Aguado-etal:98}, in which the
global PES is represented by a many-body expansion:
\begin{equation}
V_{\rm ABC} = \sum_A V_{\rm A}^{(1)} + \sum_A V_{\rm AB}^{(2)} (r_{\rm AB}) + V_{\rm ABC}^{(3)} (r_{\rm AB},r_{\rm AC},r_{\rm BC}) \,,
\end{equation}
where $V_{\rm A}^{(1)}$ represents the energy of the atoms (A$=$O,H,H) in the
ground electronic state, $V_{\rm AB}^{(2)}$ the diatomic terms (AB$=$OH,OH,HH)
and $V_{\rm ABC}^{(3)}$ the 3-body term (ABC$=$OHH). The overall rms error of
the two analytical potentials calculated over the 3500 geometries was found to
be 0.61\,kcal/mol and 0.44\,kcal/mol for the $1^3A'$ and $1^3A''$ states,
respectively.

Both PESs do not present any minimum out of the asymptotic channels, and both
show two saddle points. The first saddle-point corresponds to the reaction
barrier, with an energy 13.8\,kcal/mol above the reactant valley. It
corresponds to an O-H-H linear geometry where both states are degenerate. In
both PESs, this saddle-point was found at $r_{\rm OH}=2.28$\,a.u. and $r_{\rm
HH}=1.71$\,a.u., in good agreement with the position and energy of the saddle
point optimized at the {\em ab initio} level, which is found at an $r_{\rm OH}$
distance of 2.30\,a.u. and an $r_{\rm HH}$ distance of 1.68\,a.u, and have an
energy of 13.6\,kcal/mol above entrance channel. The second saddle-point is
found for a linear H-O-H geometry with both OH distances equal to 1.75\,a.u.,
and lies at 80.7\,kcal/mol above the entrance channel, well above the largest
energy considered in this work.

{\ }

\subsection{Dynamical Calculations}

Based on the PESs obtained, QCT and time independent QM calculations were
carried out at the three collision energies: 15\,kcal\,mol$^{-1}$,
20\,kcal\,mol$^{-1}$, and 25\,kcal\,mol$^{-1}$. QCT calculations consisted of
batches of $5 \times 10^6$ trajectories following the methodology described in
Refs.~\citenum{AHS:JCP92} and \citenum{ABH:JCSFAR98}. The trajectories were
started and finished at a atom-diatom distance of 20 a.u. ($\sim$ 10\,{\AA}),
and the integration step size was chosen to be 0.05\,fs, which guarantees a
energy conservation better than a 0.01\%. The rovibrational energy of the
reactant D$_2$ molecule was calculated by semiclassical quantization of the
action using the potential given by the asymptotic reactant valley of the PES.
The assignment of the product quantum numbers was carried out by equating the
square of the classical D$_2$ molecule rotational angular momentum to $j'
(j'+1) \hbar^2$. The vibrational quantum number $v'$ was found by equating the
internal energy of the products to a rovibrational Dunham expansion. The
``quantum numbers'' so obtained were rounded to the nearest integer.

To extract the contributions of each trajectory to the $\Pi(A')$ or $\Pi(A'')$
$\Lambda$-doublet states, it is sufficient to determine the classical product
$a^{(2)}_0$ polarization parameters with respect to the recoil direction on
each PES. This polarization parameter is given by $\langle P_2(\cos
\theta_{{\bm j}'{\bm k}'})\rangle$, where the brakets indicate the averaging
over the set of reactive trajectories leading to a given final state.

Time independent QM calculations were carried out using the ABC \cite{ABC}
code. The basis set for the calculations included all the diatomic energy
levels up to 63.4\,kcal/mol. The propagation was carried out in 150
log-derivative sectors up to a distance of 20\,a.u. For $J >0$ the value of
$\Omega_{\rm max}$, the maximum value of the projection of $J$, and the
rotational angular momentum onto the body fix axis was always chosen to be
larger than the maximum value of $j'$ energetically accessible.

\section*{Acknowledgement}
The authors acknowledge funding by the Spanish Ministry of Science and Innovation (grants
CTQ2012-37404-C02, CTQ2015-65033-P, and Consolider Ingenio 2010 CSD2009-00038). MB
gratefully acknowledges the support of the UK\ EPSRC ({\em via} Programme Grant
EP/L005913/1). PGJ acknowledges the Spanish Ministry of Economy  and Competitiveness for
the Juan de la Cierva fellowship (IJCI-2014-20615).

\section{Appendix}

\subsection{Classical and Semiclassical Deduction of Equation (5)}

\begin{figure}[ht!]
\centering
\includegraphics[width=0.9\linewidth]{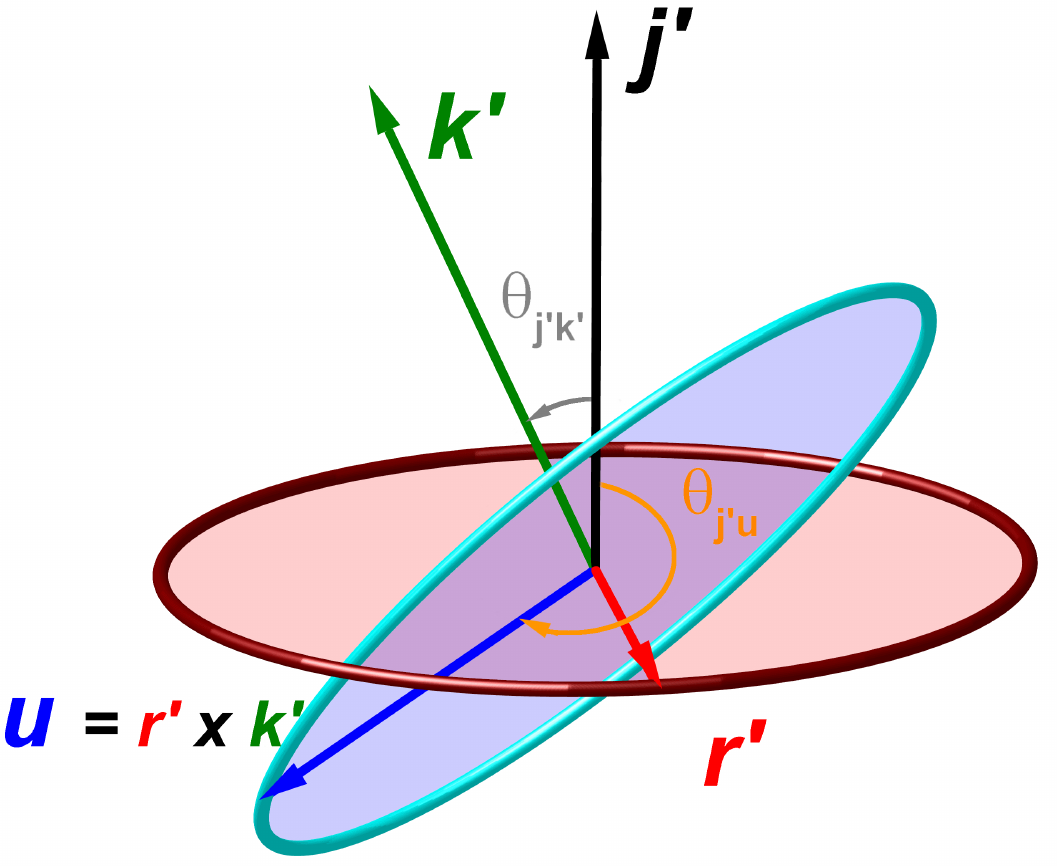}
\caption{The frame of coordinates that defines the various vectors relevant in
the quasiclassical description. The rotational angular momentum, $\bm{j'}$, is
shown in black and kept fixed along the $z$ axis. The O-H internuclear axis
($\bm{r'}$, in red) rotates perpendicular to $\bm{j'}$ and its possible values
are shaded in red. The recoil direction $\bm{k'}$, also fixed, is shown as a
green arrow. Vectors ${\bm k'}$ and ${\bm r'}$ define the three-atom plane. The
possible directions of the vector ${\bm u} = \bm{r'} \times \bm{k'}$ are shaded
in blue. For a particular $\bm{r'}$ (red vector), $\bm{u}$ is shown as a blue
arrow.} \label{frame}
\end{figure}

As pointed out in the main text, the $A'$ or $A''$ electronic symmetry of the
potential energy surfaces is defined with respect to the rotating body-fixed
DOD plane, defined by $\bm{r'}$ and $\bm{R'}$ (OD internuclear vector and the
atom-diatom D--OD vector, respectively). In turn, the symmetry of the
$\Lambda$-doublet states is defined with respect to the reflection in the OD
rotation plane that contains $\bm{r}'$ and is perpendicular to $\bm{j}'$, the
nuclear rotational angular momentum.

Therefore, the relevant angle is $\theta_{{\bm j}' {\bm u}}$; that is, the
angle between $\bm{j'}$ and $\bm{u}$, a vector in the direction of $\bm{r'}
\times \bm{R'}$.  The vector $\bm{R}'$ is asymptotically parallel to the
product recoil vector $\bm{k'}$ and hereinafter we will use the latter as
reference. In fact, $\theta_{{\bm j}' {\bm u}}$ represents the dihedral angle
between the molecular plane, and the three-atom plane. As pointed out in
Refs.~\citenum{BZ:CPL90,PHMKAB:JCP15}, $\cos^2\theta_{{\bm j}' {\bm u}}$ can be
used to relate the symmetry of the $\Lambda$-doublet levels to the symmetry of
the potential energy surface (PES).

The use of $\cos^2\theta_{{\bm j}'{\bm u}}$ stems from the fact that represents
the probability that for each PES, the OD molecule will be produced in a given
$\Lambda$-doublet state. Classically, the  use of the square of the cosine of
$\theta_{{\bm j}'{\bm u}}$ can be justified as we are interested in the mutual
alignment of the planes depicted in Fig.~\ref{frame}.

Without any loss of generality, we can select a space-fixed (scattering) frame
of coordinates in which both the product rotational angular momentum,
$\bm{j}'$, and the recoil direction, $\bm{k}'$, are fixed. Let us also assume
that $\bm{j}'$ lies along the $z$ axis (see Fig.~\ref{frame}), and $\bm{k}'$
is contained in the $xz$ plane. With this choice, the OD internuclear axis,
${\bm r}'$, will lie in the $xy$ plane. Then, $\theta_{{\bm k}'{\bm j}'}$, the
angle between $\bm{j}'$ and $\bm{k}'$, is given by:
\begin{equation}
\cos \theta_{{\bm k}'{\bm j}'} = \frac{k'_z}{ |{\bm  k'} |} \,,
\end{equation}
where $k'_z$ is the $z$ component of $\bm{k}'$. Since ${\bm u}= {\bm r}' \times
{\bm k}'$, $|{\bm u}|= |{\bm r}'| |{\bm k}'| \sin \theta_{{\bm k}'{\bm r}'}$,
where  $\theta_{{\bm k}'{\bm r}'}$ is the angle between $\bm{k}'$ and
$\bm{r}'$. Hence, the cosine of the angle between $\bm{j}'$ and $\bm{u}'$ is
given by:
\begin{eqnarray}\label{thetavj1}
\cos\theta_{{\bm j}' {\bm u}} &=& \frac{u_z}{ |{\bm u}|} = - \frac{1}{| {\bm u}
|}  r'_y k'_x  =  - \frac{| {\bm r}' | | {\bm k}'|}{|{\bm u}|} \sin\phi_{{\bm
r}'} \sin \theta_{{\bm k}' {\bm j}'} \nonumber \\ & = & - \frac{\sin\theta_{{\bm k}'{\bm
j}'}\sin\phi_{\bm r'}}{\sin\theta_{{\bm k}'{\bm r}'}} \,,
\end{eqnarray}
where $\phi_{{\bm r}'}$ is the azimuthal angle of ${\bm r}'$.

Using the law of cosines, it can be shown that:
\begin{equation}\label{cosines}
\cos\theta_{ {\bm k}' {\bm r}'} =\cos\theta_{{\bm k}'{\bm j}'}\cos\theta_{{\bm
j}'{\bm r}'} + \sin\theta_{{\bm k}'{\bm j}'}\sin\theta_{{\bm j}'{\bm r}'}\cos
\phi_{\bm r'} = \sin\theta_{{\bm k}'{\bm j}'}\,\cos\phi_{\bm r'} \,,
\end{equation}
where we have used the fact that $\bm{r}'$ is perpendicular to $\bm{j}'$.
Therefore, combining Eqs.\ \eqref{thetavj1} and \eqref{cosines} we obtain the
following expression for $\cos^2\theta_{{\bm j}' {\bm u}}$:
\begin{equation}\label{thetavj2}
\cos^2\theta_{{\bm j}' {\bm u}} = \frac{  \sin^2 \theta_{{\bm k}'{\bm j}'}
(1-\cos^2 \phi_{\bm r'})}{1 - \sin^2\theta_{{\bm k}'{\bm j}'} \,\cos^2
\phi_{\bm r'}} \,.
\end{equation}
In the chosen space-fixed reference frame, $\bm{j'}$ and $\bm{k'}$ do not
change with rotation in the product asymptote for a given trajectory. Hence,
the only variable that changes with rotation in eqn~\eqref{thetavj2} is
$\phi_{{\bm r}'}$. Averaging over a rotational period, {\em i.e.}\ integrating
eqn~\eqref{thetavj2} over $\phi_{{\bm r}'}$ and dividing by $2\pi$, the average
value of $\cos^2\theta_{{\bm j}'{\bm u}}$  over a rotational period is given
by:
\begin{equation}\label{theeq}
\langle \cos^2\theta_{{\bm j}' {\bm u}}  \rangle_{\rm rot} = 1 - |\cos^2
\theta_{{\bm k}'{\bm j}'}|^{1/2} \,.
\end{equation}

Equation~\eqref{theeq} is particularly relevant, since it relates $\theta_{{\bm
j}' {\bm u}}$, the angle between the normal vectors to the OD rotation plane
and to the three-atom plane, with $\theta_{{\bm k}' {\bm j'}}$.
Semiclassically, $|{\bm j'}| \cos\theta_{{\bm k}'{\bm j}'}$ is the projection
of ${\bm j}'$ onto ${\bm k}'$, that is, the product's helicity $\Omega'$, and
the right hand side of eqn~(5) is recovered.

\subsection{Quantum Mechanical deduction of Equation (5)}
Following refs. \citenum{ZSHA:JMS73,Alexander:JCP84,DAL:JCP89}, the OH molecular rotational wave function can be
written as
\begin{equation}
|j' \Omega' \Lambda' \rangle= \left(\frac{2j'+1}{4 \pi}\right)^{1/2} D^{j'
*}_{\Omega' \Lambda'} (\alpha, \beta, \gamma=0),
\end{equation}
where $D^{j' *}_{\Omega' \Lambda'}$ is the rotation matrix element and
($\alpha, \beta, \gamma$) are the Euler angles which specify the orientation of
the body fixed frame (BF), $xyz$, with respect to the space fixed (SF) frame,
$XYZ$. The BF frame is chosen with $z$ along the OH internuclear axis, ${\bm
r}'$, whilst the $Z$ axis in the SF frame is chosen along the recoil velocity
vector, ${\bm k}'$. With this choice, $\beta=\theta_{{\bm k}'{\bm r}'}$ and
$\alpha=\varphi_{{\bm r}'}$, which define the direction of ${\bm r}'$ in the SF
frame. The angle $\gamma$ is chosen to be zero, such that the line of nodes
(the intersection of the $XY$ and $xy$ planes) is $y \equiv {\bm u}$ which is
perpendicular to both $z={\bm r}'$ and $Z={\bm k}'$ and, as discussed in the
previous subsection, defines the normal vector to the three-body plane. The
projection of the rotational angular momentum, ${\bm j}'$, along the BF $z$
axis is $\Lambda'$. If the open shell character of the molecule is neglected,
$\Lambda'=0$. In turn, $\Omega'$ is the projection of ${\bm j}'$ along the SF $Z$ axis
(usually $m'$ is used to designate the projection of ${\bm j}'$ onto the SF
axis, but in the present case, since ${\bm k}'$  is taken as $Z$, it
corresponds to the helicity which is commonly designated by $\Omega'$). As
discussed in refs.~\citenum{Alexander:JCP84,DAL:JCP89,Alexander1988}, with this choice of frames, for $\Omega'=0$ and
$j' \gg1$, ${\bm j}'$ lies along $\bm u$ (the $y$ axis) and $xz$ is the rotation
plane. For $\Omega'=j'$ and $j' \gg 1$, ${\bm j}'$ is along the $x$ axis, which in
this case is along $-Z$ and the rotation plane is $yz$.

Classically, $|{\bm j'}|^2 \cos^2\theta_{{\bm j}' {\bm u}}$ represents the
square of the projection of ${\bm j'}$ onto the $\bm u$ vector. In QM, the
equivalent magnitude would be $\langle \hat{j'}_{\bm u}^2 \rangle$, the
expectation value of the square of the operator that represents the projection
along ${\bm u}$. It can be shown that the expression of $\hat{j'}_{\bm u}$
is simply $-i \partial/{\partial \theta_{{\bm k}'{\bm r}'}}$. \cite{Zare} Therefore
\begin{eqnarray}
&&\langle j'^{\,2}_{\bm u} \rangle= \langle j' \Omega' \Lambda'|\hat{j'}_{\bm
u}^2|j' \Omega' \Lambda' \rangle= \nonumber  \\ &&\frac{2j'+1}{4
\pi} \int_0^{2\pi} \!\! \int_{-1}^{1}   D^{j'}_{\Omega' \Lambda'} (\varphi_{{\bm
r}'},\theta_{{\bm k}'{\bm r}'} , 0) \nonumber \\ && \cdot \left(-\partial^2/{\partial \theta^2_{{\bm
k}'{\bm r}'}} \right) D^{j'*}_{\Omega' \Lambda'} (\varphi_{{\bm
r}'},\theta_{{\bm k}'{\bm r}'} , 0) {\rm d}\varphi_{{\bm r}'} {\rm d}\cos
\theta_{{\bm k}' {\bm r}'}
\end{eqnarray}
It can be shown that the result of this integral is almost exactly
\begin{eqnarray}
&&\langle j'^{\,2}_{\bm u} \rangle= j'(j'+1-\delta_{\Omega',0}) \left( 1- \left|
\frac{\Omega'^2}{j'(j'+1)} \right|^{1/2} \right)
\end{eqnarray}
where  $\delta_{\Omega',0}$ stems from the fact that for $\Omega'$=0 the
maximum value of the projection is $j'$. Apart from this correction, this
equation is the semiclassical expression, eqn (5) of the main text.


\balance


\bibliographystyle{rsc} 

\providecommand*{\mcitethebibliography}{\thebibliography}
\csname @ifundefined\endcsname{endmcitethebibliography}
{\let\endmcitethebibliography\endthebibliography}{}

\end{document}